\documentclass[aps,prb,twocolumn]{revtex4-1}
\usepackage{graphicx} 
\usepackage{amsmath}
\newcommand*{\citen}[1]{%
  \begingroup
    \romannumeral-`\x 
    \setcitestyle{numbers}%
    \cite{#1}%
  \endgroup   
}
\begin{document}
\title{CDW-Exciton condensate competition and a condensate driven force} 
\author{E. \"{O}zg\"{u}n$^{\bf (1), \dagger}$ and T. Hakio\u{g}lu$^{\bf (2)}$} 
\affiliation{
${\bf (1)}$ {Department of Physics, Bilkent University, 06800 Ankara, Turkey}
\break
${\bf (2)}$ {Consortium of Quantum Technologies in Energy (Q-TECH), Energy Institute, \.{I}stanbul Technical University, 34469, \.{I}stanbul, Turkey}
\break
${\dagger}$ {Nanotechnology Research Center, Bilkent University, 06800 Ankara, Turkey (Current Address)}}

\begin{abstract}
We examine the competition between the charge-density wave (CDW) instability and the excitonic condensate (EC) in spatially separated layers of electrons and holes. The CDW and the EC order parameters (OPs), described by two different mechanisms and hence two different transition temperatures  $T^{CDW}_c$ and $T^{EC}_c$, are self-consistently coupled by a microscopic mean field theory. We discuss the results in our model specifically focusing on the transition-metal dichalcogenides which are considered as the most typical examples of strongly coupled CDW/EC systems with atomic layer separations where the electronic energy scales are large with the critical temperatures in the range $T^{EC}_c \sim T^{CDW}_c \sim 100-200 K$. An important consequence of this is that the excitonic energy gap, hence the condensed free energy, vary with the layer separation resulting in a new type of force ${\cal F}_{EC}$. We discuss the possibility of this force as the possible driver of the structural lattice deformation observed in some TMDCs with a particular attention on the $ 1 {\it T}$-$TiSe_2$ below $200 K$. \end{abstract}

\maketitle         
\section{Introduction} 
In this article we examine the competition between the CDW formation and the EC in layered systems. Long time ago Balseiro and Falicov (BF) developed a model based on the competing orders of SC and CDW formation.\cite{BF} The model quickly became important in the formulation of pseudogap in the SC of the cuprates. Even earlier, the CDW was shown to arise when the Fermi level is pinned to a Van Hove singularity.\cite{TMR_GKS}

On the other hand, the transition-metal dichalcogenides (TMDCs) are considered to be typical examples where a number of low temperature phases can be mutually coupled. Theoretical studies on the optical, electronical and structural properties in bulk or layered TMDCs showed the coexistence of the structural instability, antiferromagnetism and conventional superconductivity (SC)\cite{WY}. Their electronic properties are rich with insulating, semiconducting, semi-metal or true metal behaviour. Many layered TMDCs also show excitonic superconductivity coupled with lattice distortion. There, the excitonic coupling can be much stronger than ordinary semiconductors due to the high electronic energy scales of $\sim 100 meV$. In  Ref.\citen{WY}, the coexistence of the CDW and SC was speculated, yet reliable experiments came much later. A detailed review of more recent experimental and theoretical progress can be found in Ref.\citen{GVA}. In this work, we are motivated by the BF model to study the CDW/EC coupled phase in TMDCs. 

EC is analogous to SC with an exception of the charge neutrality of the pairs. The Coulomb pairing is the most fundamental interaction in EC. In the materials of interest here, this energy scale is much higher than the phonon mediated pairing energy in conventional SC. Based on these, we present in this work, a scenario for the CDW/EC ordered systems where the CDW is driven by a strong phonon coupling separately in the electron and hole layers whereas the EC is driven electronically by the Coulomb interaction between these layers. Hence two different mechanisms lead to two critical temperatures $ T^{CDW}_c $ and $ T^{EC}_c $ which can have, due to their self-consistent coupling, unequal but closely related scales. The ordering between the two temperatures is dictated by the phonon and the electronic energy scales. For instance, it was recently argued in the context of the specific TMDC material $ 1 {\it T}$-$TiSe_2$ that\cite{Monney4} $T^{CDW}_c > T^{EC}_c$ where both temperatures are within the $ 100-200K $ range. More recently $T^{EC}_c > T^{CDW}_c$ is also being discussed since EC is believed to be the precursor of a low temperature lattice deformation. 

In order to motivate the reader to the model devised in the next section, a brief discussion about the TMDCs is needed. These are given by $MX_2$, where $M$ is a transition metal and $X$ is a chalcogen atom, which are layered compounds consisting of three stacking layers within the unit cell where $M$ atoms, located in the middle layer, are sandwiched between the two $X$-like layers. In the mid 70's, superlattice formation was reported\cite{Woo, Salvo} in experiments with $ 1 {\it T}$-$TiSe_2$. Theoretical works appeared much later searching for a microscopic mechanism behind this instability\cite{Suzuki} which still remains contraversial. Three scenarios are on debate: a) Fermi surface (FS) nesting, b) Jahn-Teller effect, c) excitonic condensation. While the former is reported as a weak candidate, latter two are supported by experiments. ARPES studies in favor of Jahn-Teller scenario\cite{Rossnagel}, those supporting excitonic insulator scenario\cite{Pillo} and experiments supporting both scenarios\cite{Kidd} were reported. Although numerous other experimental results exist, the most recent studies point at the excitonic insulator scenario\cite{Monney1, Monney2, Monney3}, which is further supported by the relatively high value of the transition temperature $ T^{EC}_c $. In these layered compounds, it is established that\cite{Monney2}, excitonic effects arise from the conduction bands dominated by the $3d$ even-parity states of the relevant $M$ orbitals and the valence bands mainly in the $4p$ odd-parity $X$ orbitals. The nearly fixed even and odd parities of the bands respectively of $X$-type 4p and $M$-type 3d character imply that the parity mixing and the $M-X$ hybridization is weak in these bands. This is shown to be the result of the octahedral coordination splitting the $3d$-like conduction bands and opening a van der Waals gap.\cite{vanwezel} 
  
A typical monolayer thickness\cite{10A} in TMDCs is on the order of $6.5 \,\AA$ hence the $M$-$X$ separation can be approximately taken as $3$-$4 \,\AA$. The weak hybridization in the excitonic sector creates a natural formation of coupled electron-hole quantum wells (EHQW). Based on this, it is proposed in Ref.\citen{Monney3} that the CDW and the periodic lattice distortion are created by the formation of an EC\cite{Salvo} where excitons are coupled to the CDW phonons through a Fr\"ohlich type interaction. The critical temperatures were believed to be equal, i.e. $T^{EC}_c=T^{CDW}_c$ and in the $100-200$K range due to the large electronic energy scales. The recent experiments based on monolayer samples have discovered the same CDW/EC transition as in the bulk with the same $T_c$, stressing the two dimensional character of the transition and ruling out any scenario connected with the third dimension.\cite{Sugawara} These new findings support that these materials are superior natural candidates for excitonic EHQW. Additionally, the two-dimensional natural EHQW geometry is experimentally important, since EC was observed recently after long years of search in artificially grown EHQWs.\cite{ECbutov} 

A recent attempt to investigate the origin of the periodic lattice distortions in the specific TMDC $1T$-$TiSe_2$ uses variational Monte Carlo method to solve the two band Hubbard model in a triangular lattice\cite{JAP}. The interband and intraband Coulomb interactions were replaced by the respective hardcore interaction constants $U$ and $U^{\prime}$. The momentum dependence of the Coulomb interaction however plays a significant role in the momentum dependent order parameter of the EC and its dependence on the electron-hole layer separation $D$ should naturally be incorporated in any realistic model. The condensed free energy of the CDW/EC phase is then a function of $D$. Hence an internal stress should be expected to built up due to the condensation. We propose in this work that, this stress, which we call as the {\it EC-force} ${\cal F}_{EC}$, generates a uniform strain field which can explain the observed lattice deformation. On the other hand, the intralayer Coulomb interaction can be destructive on the nesting properties of the Fermi surface which can suppress or even destroy the intralayer CDW formation. Therefore any mechanism related to the CDW formation should properly account for its presence. These intralayer mechanisms are however, independent of $D$ and their direct energetic influence on the EC force is negligible, as we do in this work (shown below). The effect of imperfect CDW nesting, on the other hand, can still be modelled through the next-nearest-neighbour interaction $t_1$ as we do in this work.   


The concept of a force arising from condensation is not new to this work. The EC-force was first predicted in numerical\cite{SSC} and semi-analytical\cite{APL} calculations in EHQW heterostructures of III-V semiconductor compounds, where the relevant energies, i.e. the Hartree energy $E_H$ and the Fermi energy $E_F$ are both in the $10 meV$ range. However, the CDW and EC interaction strengths, the hopping energy scale $t_0\simeq 100-200 meV$ and the critical temperatures of $150-200 K$ in the TMDC materials are at least an order of magnitude larger than the system studied in Ref's\citen{SSC, APL}. Also the exciton Bohr radius\cite{Zhang} $ a_B^{TMDC}\simeq 8-10\AA$ is nearly one order of magnitude smaller than the semiconductor $a_B^{EHQW}\sim 100 \AA$. We hence expect that this internal stress can be much stronger here, which can make them important candidates in the experimental search for ${\cal F}_{EC}$. Here we demonstrate that, a structural deformation of $(1-10)\times 10^{-3} \AA$ can be accounted for by ${\cal F}_{EC}$ which is in the same order of magnitude as reported in the experiments\cite{Salvo}. This opens a possibility that the lattice distortion observed in the TMDCs indicates the emergence of the {\it EC-force}. We now devise a microscopic model to quantify this effect.  

\begin{figure} 
\includegraphics[scale=0.7,angle=0]{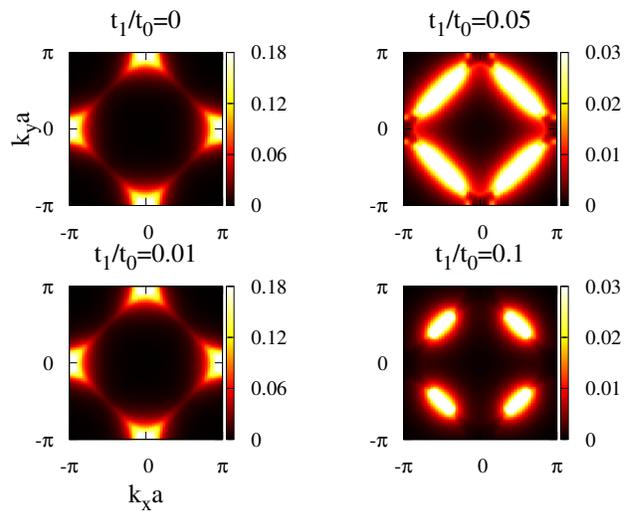}
\vspace*{0.5cm}
\caption{(Color online) EC OP, scaled with $ t_0=0.125 eV$, is plotted for different second NN interaction strengths. The peak positions of the EC OP are separated by the nesting vector, ${\bf Q} = (\pm\pi,\pm\pi)  $ in each of the four cases. For zero or a small second NN interaction, OP is maximum at the saddle points of the dispersion, due to nearly perfect nesting. As the second NN interaction increases, the perfect nesting gradually disappears. As the result, the energy is strongly lowered by the EC OP within small pockets at points not coinciding with a reciprocal lattice vector (lower right plot).}
\label{delta}
\end{figure}

\begin{figure}
\vspace*{1cm}
\hspace{-1.8cm} 
\includegraphics[scale=0.7,angle=0]{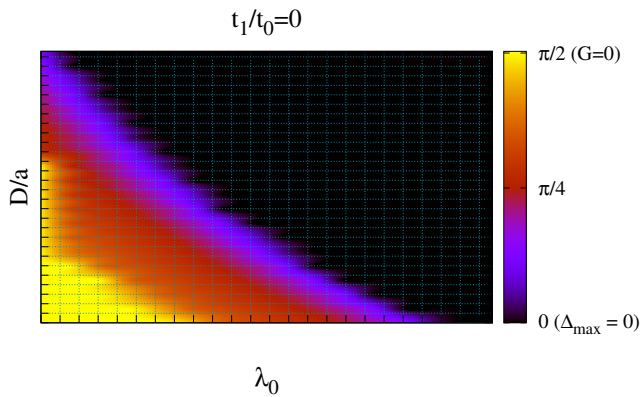}
\caption{(Color online) Color map of the CDW and EC for $ t_1= 0 $. The OPs are mapped via $ f_{col} = tan^{-1}\big [ \frac{\Delta_{max}}{G}  \big ] $ transformation. In yellow (light) regions there is only EC and in black (dark) regions only CDW is present, whereas in between they coexist. Here, $ \lambda_0 $ runs from $ 0.9 $ to $ 1.6 $ and $ D/a $ varies between $ 2 $ and $ 4 $.}  
\label{phase}
\end{figure}
\section{Microscopic theory of the CDW/EC system} 
Our microscopic approach is based on the CDW/EC competition in which we extend the BF formalism in Ref.[\citen{BF}] to the 2D coupled EHQW geometry. Our goal is to solve the coupled CDW/EC system for an order of magnitude estimation of the EC-force and for studying its influence by  the CDW background. In doing this, we simplify the picture where we assume one electron and one hole bands with symmetric properties (equal effective mass, equal chemical potential) spatially confined within separate layers. In relation to TMDC, this would imply that the weak hybridization between the $M$ and the $X$ like bands is ignored. 
We will also assume a square lattice for both layers since the EC-force is a condensation phenomenon in the bulk which does not crucially depend on the lattice structure. A strong FS nesting within each layer is also present in the model which can drive a CDW order and in addition, the two layers are coupled electronically by a short range attractive Coulomb interaction. It is crucial to note that, although a nesting driven CDW is a part of the experimental reality, the  ${\cal F}_{EC}$ and the lattice deformation arising from the EC does not crucially depend upon the CDW formation. We propose in this work that, the lattice deformation is a static strain field which is induced by the internally built-in stress once the EC is formed. Here it is therefore implicit in our model that the Fermi surface nesting and the crystal symmetry are not crucial in the mechanism leading to lattice deformation and our results are equally valid in their absence (see below). 

CDW formed in the "upper", lets say $X$ layer is,   
\begin{eqnarray}
\langle \hat{\rho}_{u}{(\bf r)}\rangle = n_0+ \gamma_u n_1 cos({\bf Q}\cdot{\bf r})
\label{den}
\end{eqnarray}
where $ \hat{\rho}_{u}{(\bf r)} =\hat{u}^{\dag}({\bf r})\hat{u}({\bf r}) $ is the density operator of the upper layer with $ \hat{u}^{\dag}({\bf r})/\hat{u}({\bf r}) $ being creation/annihilation operators in real space, $ n_0 $ is the mean density, $n_1$ is the CDW amplitude, $ {\bf Q} = (\pi,\pi) $ is the nesting vector satisfying $ {\bf k} + 2{\bf Q} = {\bf k} $ and $ \gamma_u=1 $. There is a similar expression for the "down", i.e. the $M$, layer with $ \hat{d}^{\dag}({\bf r})/\hat{d}({\bf r}) $ and $ \gamma_d=-1 $. Hence, the CDWs in both layers built a relative $ \pi $-shift to avoid the strong Coulomb repulsion. The interlayer Coulomb interaction is given by  
\begin{eqnarray}
 \hat{v}_{int} = \int d{\bf r} d{\bf r^{\prime}} \hat{\rho}_u({\bf r})V({\bf r}-{\bf r^{\prime}})\hat{\rho}_d({\bf r^{\prime}})
\label{Coulomb_int_1}
\end{eqnarray} 
where the Coulomb coupling $V({\bf r}-{\bf r^{\prime}})=e^2/(4\pi\epsilon\vert {\bf r}-{\bf r^\prime}-D {\bf e}_z\vert)$ with $e, \epsilon$, ${\bf e}_z$ and $D$ as the electric charge, the dielectric constant, unit vector in $z$-direction and the layer separation. Using Eq.(\ref{den}) in Eq.(\ref{Coulomb_int_1}), we find three terms: a) a repulsive contribution proportional to $n_0^2$, b) two contributions linear in $n_0$ that cancel out and, c) a term proportional to $n_1^2$. The mean field of Eq.(\ref{Coulomb_int_1}) is,  
\begin{eqnarray}
\langle \hat{v}_{int} \rangle =  \int d{\bf r} d{\bf r^{\prime}} V({\bf r}-{\bf r^{\prime}}) [n_0^2-\tilde{n}_1({\bf r}) \tilde{n}_1({\bf r^{\prime}})]
\label{Coulomb_int_2}
\end{eqnarray}  
where $ \tilde{n}_1({\bf r}) = n_1 cos({\bf Q}\cdot{\bf r}) $. The repulsive first term is a constant which can be absorbed into the chemical potential. The second term is attractive due to the   $\pi$-shift, and if $ n_1 \neq0 $, the layers are coupled as electron-hole layers. This interaction creates an instability at a critical strength which opens an excitonic gap in the spectrum. We use a simple tight-binding model that can reveal the CDW/EC competition in a square lattice geometry and a band dispersion of, $\epsilon_{{\bf k}} = -2t_0[cos(k_xa)+cos(k_ya)]-4t_1cos(k_xa)cos(k_ya)$, where $ t_0 $ and $ t_1 $ are the first and the second nearest neighbour (NN) interactions. 
Here $t_1$ is introduced as a measure of the degree of nesting, i.e. $t_1=0$ for perfect and $t_1 >> 0$ for weak nesting whereas $t_0$ is connected with the bandwidth which determines the critical temperature. We use a $t_0$ range such that the relevant $T_c$'s are within the $100-200$ K range. Considering the CDW in Eq.(\ref{den}), $n_0 = [1/(2\pi)^2]\int {\bf dk} \langle \hat{u}^{\dag}_{\bf{k}}\hat{u}_{\bf{k}} \rangle$ and $n_1=G/2 \lambda_{ep}$ with $\langle \hat{u}^{\dag}_{\bf{k}}\hat{u}_{\bf{k}} \rangle=\langle \hat{d}^{\dag}_{\bf{k}}\hat{d}_{\bf{k}} \rangle$ are the important correlations\cite{BF}, where $\lambda_{ep}$ is electron-phonon coupling strength and,
\begin{eqnarray}
G = \begin{cases} (G_0+G_1) \quad {\rm if} \quad |\epsilon_{\bf{k}}-\mu| < \hbar w_D \cr G_0 \quad {\rm otherwise} \end{cases} 
\label{conditional}
\end{eqnarray} 
Here, $\lambda_{ep}$ is assumed to be approximately k-independent within a Debye energy range and $ \mu $ is the chemical potential. The CDW order parameters are 
\begin{eqnarray}
G_0 &=& \lambda_{ep} \int^{\prime\prime} \frac{{\bf dk}}{(2\pi)^2}\langle \hat{u}^{\dag}_{\bf{k}}\hat{u}_{\bf{k}+\bf{Q}} \rangle \qquad {\rm and}
\label{G_zero} \\
G_1 &=& \lambda_{ep} \int \frac{{\bf dk}}{(2\pi)^2}\langle \hat{u}^{\dag}_{\bf{k}}\hat{u}_{\bf{k}+\bf{Q}} \rangle 
\label{G_one}
\end{eqnarray}
In the above equation, the double primed integral is performed only when $|\epsilon_{\bf{k}}-\mu| < \hbar w_D$ and $|\epsilon_{\bf{k}+\bf{Q}}-\mu| < \hbar w_D$ hold.\cite{BF} We have self consistency conditions: $ \langle \hat{u}^{\dag}_{\bf{k}}\hat{u}_{\bf{k}} \rangle =  \langle \hat{d}^{\dag}_{\bf{k}}\hat{d}_{\bf{k}} \rangle  $ and $ \langle \hat{u}^{\dag}_{\bf{k}}\hat{u}_{\bf{k}+\bf{Q}} \rangle = - \langle \hat{d}^{\dag}_{\bf{k}}\hat{d}_{\bf{k}+\bf{Q}} \rangle $. The first condition comes from our assumption that the layers are identical, and the number of particles within each layer are the same. The second one is due to the $\pi$-shift between the CDWs within the individual layers. In the model we consider, the CDW and the EC OPs are coupled self consistently. While the former is driven by $\lambda_{ep}$, as formulated in Eqs.(\ref{conditional}-\ref{G_one}), the latter is driven by the interlayer Coulomb interaction in Eq.(\ref{Coulomb_int_2}). Using the Hartree-Fock mean field approximation, the  Hamiltonian is given in the $(\hat{u}^{\dag}_{{\bf k}} \, \hat{u}^{\dag}_{{\bf k + Q}} \, \hat{d}^{\dag}_{{\bf k}} \hat{d}^{\dag}_{{\bf k + Q}})$ basis by, 
\begin{eqnarray}
{\cal H}= \sum_{\bf k} \Bigg \{ H_0 + \begin{pmatrix} \epsilon^{(-)}_{{\bf k}} & -G & \Delta^{(1)}_{{\bf k}} & \Delta^{(2)}_{{\bf k}} \cr -G & -\epsilon^{(-)}_{{\bf k}} & \Delta^{(2)}_{{\bf k}} & \Delta^{(1)}_{{\bf k + Q}} \cr \Delta^{(1)}_{{\bf k}} & \Delta^{(2)}_{{\bf k}} & \epsilon^{(-)}_{{\bf k}} & G \cr \Delta^{(2)}_{{\bf k}} & \Delta^{(1)}_{{\bf k + Q}} & G & -\epsilon^{(-)}_{{\bf k}} \end{pmatrix} \Bigg\}
\label{Ham}
\end{eqnarray}
where the spin degrees of freedom of the EC OPs\cite{PRL1, PRL2} are eliminated due to the spin degeneracy. Furthermore, dark versus bright exciton difference \cite{Comb,SSC} is out of the scope of this manuscript and omitted here. In Eq.(\ref{Ham}), $H_0 = (\epsilon^{(+)}_{{\bf k}}-\mu){\bf \sigma_0}\otimes{\bf \sigma_0} $, $ \epsilon^{(+)}_{{\bf k}} = (\epsilon_{{\bf k}}+\epsilon_{{\bf k+Q}})/2$ , $ \epsilon^{(-)}_{{\bf k}} = (\epsilon_{{\bf k}}-\epsilon_{{\bf k+Q}})/2 $ and $ \sigma_0 $ is the $ 2\times 2 $ unit matrix. The excitonic part has two different type of pairings, i.e. $\langle \hat{u}^{\dag}_{{\bf k}} \,\hat{d}_{{\bf k}}\rangle $ and $\langle \hat{u}^{\dag}_{{\bf k}} \,\hat{d}_{{\bf k+Q}}\rangle $, which we denote by $\Delta^{(1)}_{\bf k}$ and $ \Delta^{(2)}_{\bf k}$ respectively. In the presence of strong CDW, $ \Delta^{(2)}_{\bf k}$ dominates the ground state. We hence assume that $ \Delta^{(2)}_{{\bf k}}\ne 0$ and $ \Delta^{(1)}_{{\bf k}}$ and $\Delta^{(1)}_{{\bf k + Q}}$ are negligible. In the case of weak or vanishing CDW, $ \Delta^{(1)}_{{\bf k}}$ will be dominant in the ground state with a magnitude comparable to $ \Delta^{(2)}_{{\bf k}}$ of the strong CDW phase. 

At this point we redefine $\Delta^{(2)}_{{\bf k}}$ as $\Delta_{{\bf k}}$ given as, $\Delta_{\bf k} = (1/2A) \sum_{{\bf k^{\prime}}} v_{eh}({\bf k-k^{\prime}})  \langle \hat{u}^{\dag}_{\bf{k^{\prime}+Q}}\hat{d}_{\bf{k^{\prime}}} \rangle $, where $ A $ is the sample area, $ v_{eh}({\bf k-k^{\prime}}) = -e^2 e^{-|{\bf k-k^{\prime}}|D}/(2\varepsilon |{\bf k-k^{\prime}}|) $ is the Fourier transform of the interlayer pairing interaction $V({\bf r}-{\bf r^\prime})$. The energy spectrum is two-fold degenerate and given by, $E_1 = E_0 + \Lambda $, $E_2 = E_0 - \Lambda $ with $E_0 = \epsilon^{(+)}_{{\bf k}}-\mu $ and $\Lambda = [(\epsilon^{(-)}_{{\bf k}})^2+\Delta_{hyb}^2]^{1/2}$ where $\Delta_{hyb}=(G^2+\Delta^2_{\bf k})^{1/2}$ is the hybrid CDW/EC gap. Final expressions for the OPs to be solved numerically are,
\begin{eqnarray}
G_0 &=& -\lambda_{ep}(G_0+G_1) \int^{\prime \prime} \frac{\bf{dk}}{(2\pi)^2} \frac{F({\bf k})}{2\Lambda} \label{scs1} \\
G_1 &=& -\lambda_{ep} \int \frac{\bf{dk}}{(2\pi)^2} \frac{G}{2\Lambda}F({\bf k}) \label{scs2}  \\
\Delta_{\bf k} &=& -\frac{1}{2}\int \frac{\bf{dk^{\prime}}}{(2\pi)^2}\frac{e^2}{2\varepsilon}\frac{e^{-|{\bf k- k^{\prime}}|D}}{|{\bf k- k^{\prime}}|} \frac{\Delta_{\bf k^{\prime}}}{2\Lambda}F({\bf k^{\prime}})
\label{scs3}
\end{eqnarray}
where the first two are obtained from Eqs.(\ref{G_zero}) and (\ref{G_one}) in the ground state of the Hamiltonian in Eq.(\ref{Ham}). Here $F({\bf k})=f_1({\bf k})-f_2({\bf k})$ with $ f_\nu({\bf k}) $ with $\nu=1,2$ are the Fermi-Dirac distributions $f_{\nu}({\bf k})=1/[1+exp(\beta E_\nu)]$ where $ \beta =1/k_B T $, $ k_B $ is the Boltzmann constant and $ T $ is the temperature. Due to the opening of the hybrid gap, the condensation free energy is lowered. The $D$ dependence in $\Delta_{hyb}$ has two sources: a) when $\partial \Delta_{\bf k}/\partial D\ne 0$; which is the direct manifestation of the condensation, b) when $\partial G/\partial D\ne 0$. In the absence of EC, the CDW ordering is a completely intralayer phenomenon and the free energy is independent of D. But when EC is present, one has to carefully consider the competition between the two. It is known that SC and CDW tend to weaken each other\cite{BF} when both are driven by the same interaction. A similar conclusion is obtained here although two different mechanisms are present for the CDW and the EC. Thus both OPs contribute to the EC force as,
\begin{eqnarray}
{\cal F}_{EC} = -\frac{\partial \Delta\Omega}{\partial \Delta_{hyb}} \frac{\partial \Delta_{hyb}}{\partial D}
\end{eqnarray} \\
here $ \Delta\Omega = \Omega_O-\Omega_N$ is the free energy difference between the CDW/EC ordered and normal states respectively where\cite{APL},
\begin{eqnarray}
\Omega_O &=& -A\frac{G_0 G_1}{\lambda_{ep}} + \sum_{{\bf k}}\bigg [ \frac{\Delta^2_{\bf k}}{2\Lambda}F({\bf k}) \nonumber \\ &+& \frac{\partial}{\partial\beta}\sum_{\nu} ln(1-f_{\nu}({\bf k})) \bigg] \\ 
\Omega_N &=& \frac{\partial}{\partial\beta}\sum_{{\bf k},\nu}ln(1-f_{\nu\,0}({\bf k})) \nonumber
\end{eqnarray} 
and $f_{\nu\,0}({\bf k})$ is the Fermi-Dirac distribution $f_{\nu}({\bf k})$ when $\Delta_{hyb}=0$. We can now show that, once the hybrid gap is formed, ${\cal F}_{EC}$ is turned on which can induce a strong lattice deformation. 

\begin{figure}
\vspace*{1cm}
\hspace{-0.7cm} 
\includegraphics[scale=0.7,angle=0]{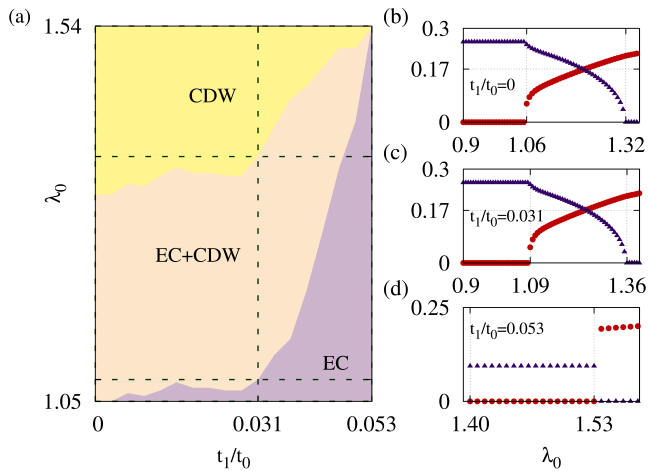}
\caption{(Color online) Regimes with different coexistence/competition properties are presented for EC and the CDW OPs for varying $ \lambda_{0} $ and $t_1$. Here, increasing $t_1/t_0$ plays the major role in breaking the optimal nesting condition which weakens both OPs, whereas $t_1/t_0$ and $\lambda_{0}$ together determine two regimes of coexistence/competition as indicated in (a). Several cross sections of (a) are given for the EC and CDW order parameters as, b) $t_1=0$: EC OP (blue triangles) gradually drops to zero with the onset of CDW (red circles), c) $t_1=0.031 t_0$: the region of coexistence is shifted to higher $\lambda$ values, and d) $t_1=0.053 t_0$: a direct transition from EC to CDW, with no coexistence. The OPs on the vertical scale of (b-d) are given in units of $ t_0 $.} 
\label{regions}
\end{figure}

We can estimate the amount of deformation by assuming harmonic conditions given by the axial stiffness coefficient $k=AE/D $ where $ A $ is the cross-sectional area and $ E $ is the Young's Modulus. The local deformation created by the local stress is  
\begin{eqnarray}
\Delta x = \frac{{\cal F}_{EC}}{k} = \frac{{\cal F}_{EC} D}{AE}~.
\label{dist}
\end{eqnarray} 
Eq.(\ref{dist}) is correctly independent from the sample size. We now find the solution of the self-consistent model in Eq.'s(\ref{scs1}), (\ref{scs2}) and (\ref{scs3}) and examine the competition between the CDW and the EC.   
       
\section{The CDW/EC system} 

In Fig.\ref{delta} the EC OP $ \Delta_{\bf k} $ is calculated for various $ t_1 $ values, in units of $ t_0 $. When $ t_1$ is small, we have nearly perfect nesting. 
In this calculation, the chemical potential is fixed at $\mu=0$ coinciding at $t_1=0$ with the nesting singularity in the density of states. The maxima of $ \Delta_{\bf k} $ are connected by ${\bf Q} = (\pm\pi,\pm\pi)  $. As $ t_1 $ is increased, perfect nesting is destroyed. As a result $  \Delta_{\bf k} $ is weakened and shifts towards the center. The competition between the EC and CDW OPs is demonstrated in Fig.\ref{phase} as $ D/a$ and the dimensionless electron-phonon coupling constant $\lambda_0= \lambda_{ep}/(a^2t_0)$ are varied for $t_1=0$. The numerical values of $\lambda_{ep}$ and $t_0$ are chosen such that the $T_c^{EC}$ and $T_c^{CDW}$ are within the 100-200 K range. The yellow (light) regions designate the pure EC and the dark regions represent pure CDW. In between they coexist and compete.    

On the other hand, when $t_1\ne 0$ the maximal nesting condition is broken. In the $t_1, \lambda_{0}$ space two different regimes of coexistence are observed as illustrated in (Fig.{\ref{regions}}.(a)) and its cross sections (Fig.{\ref{regions}}.(b-d))  : i) For $ t_1=0$ or small, the EC can coexist with the CDW in a narrow region of $\lambda_{0}$ as shown in Fig.{\ref{regions}}.(b,c). ii) For higher $t_1$ the CDW and EC OPs exclude each other completely as indicated by Fig. {\ref{regions}}.(d). When $t_1$ is further increased, the perfect nesting is strongly broken, hence the region of coexistence becomes narrower and the critical $\lambda_{0}$ is shifted to higher values as shown in the phase diagram Fig.{\ref{regions}}.(a).
\begin{figure}
\vspace*{1.2cm}
\hspace{-1.8cm} 
\includegraphics[scale=0.7,angle=0]{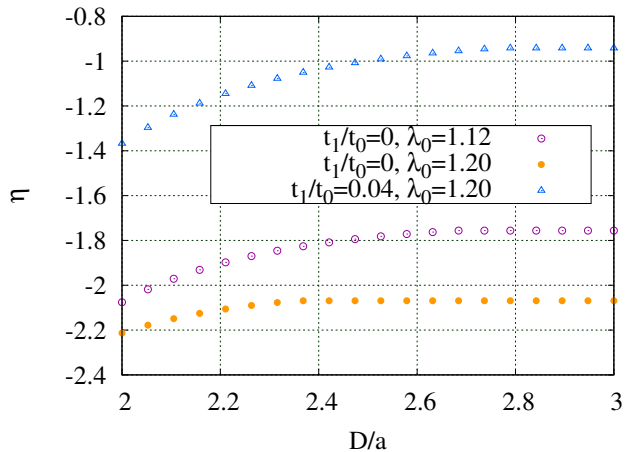}
\caption{(Color online) The change in the free energy per area with respect to $D/a$ is plotted for different $ \lambda_0 $ and $ t_1/t_0 $ values. Free energy becomes constant after EC vanishes, with only CDW remaining, which means that the EC force is zero beyond that critical point.} 
\label{free}
\end{figure}

\section{EC-force driving the lattice deformations} 
We are now in a position to report the emergence of a large EC force in this coupled CDW/EC system. The change in the free energy is shown in Fig. \ref{free} for various $ t_1/t_0 $ and $ \lambda_0 $ values. Flattening of the curves in the figure points at an important difference between this CDW/EC system and the pure EC system investigated in Ref.\citen{APL}. Here the EC force weakens as the critical point is approached, whereas in the pure EC system in Ref.\citen{APL}, the EC force is strongest at the critical point. The difference is due to the CDW in the current system, which smoothens the free energy as the EC gets weaker. 

We previously calculated the EC pressure, i.e. the EC-force ${\cal F}_{EC}$ per unit area, in AlGaAs EHQW semiconductors\cite{APL} and predicted a magnitude on the order of $1 Pa$. When the current theory is applied to a typical TMDC we find $\approx 10^7 Pa $. This enormous difference is expected due to the much stronger interactions in the current system compared to the III - V semiconductors, but still needs a qualitative explanation. The EC pressure is,   
\begin{eqnarray}
\frac{{\cal F}_{EC}}{A} =-\alpha a \frac{\partial \eta}{\partial D},\qquad  \alpha=\frac{t_0}{a^3}~, \qquad \eta =\frac{\Delta \Omega}{t_0}\frac{a^2}{A}
\end{eqnarray} 
where $\alpha$ has the pressure unit and $\eta$ is the dimensionless change of free energy per area plotted in Fig.~\ref{free}. Compared to the EHQW case, our length scale $ a $ is $ 20 $ times smaller and energy scale $ t_0 $ is about $ 10 $ times larger. So five orders of magnitude comes from $\alpha$. On the other hand, the condensation free energies in both systems have $2$ orders of magnitude difference, since in the pure EC system in semiconductor EHQWs the critical temperature is within the $1-5$ K range whereas here in the CDW/EC system it is in the $100-200 K$ range. 

We are at the point to examine the lattice deformation and propose that this enormous internal stress created by ${\cal F}_{EC}$ can be its driving mechanism of the lattice instability in many TMDC materials. We adopt $ 1 {\it T}$-$TiSe_2$ as the typical TMDC to compare our results. Although this material has a trigonal crystal symmetry, as opposed to the tetragonal one here, crystal structure is unimportant in the completely condensate driven EC-force. Also, the electrons and holes are not unstably formed in a direct band structure like in III-IV semiconductors. The presence of excitons is rather an equilibrium property observed between the individual electron-like and the hole-like pockets of the Brillouin zone. It has been established by comparing experimental features in monolayer as well as bulk samples that\cite{Sugawara} the Fermi surface nesting is ruled out as a likely cause of the lattice instability in $ 1 {\it T}$-$TiSe_2$. This consequently puts the lattice symmetry at a minor importance among other factors affecting the instability studied here. 

The lattice deformation is predicted by Eq.(\ref{dist}) using $E_{Ti} \simeq 100 GPa$ for the Young's modulus\cite{titanbook}. We find $ \Delta x \simeq (1-10)\times 10^{-3} \AA $ which, in this particular case corresponds to the change in the $Ti-Se$ distance. This result is quite agreeably compared with the experimental observations\cite{Salvo}. This not only justifies the large magnitude we found for ${\cal F}_{EC}$ but also suggests a intriguing scenario for the lattice distortions observed in TMDCs, in particular $ 1 {\it T}$-$TiSe_2$. The lattice deformation is therefore pinned to the formation of the EC and the critical temperature $T_c^{EC}$.    

The possibility of the electron-hole coupling in the periodic lattice distortions and the presence of strong excitonic background was suggested in the experimental work of Di Salvo {\it et al.}\cite{Salvo} for the typical TMDC material $ 1 {\it T}$-$TiSe_2$. Based on this, Monney {\it et al.} in Ref.\citen{Monney3} suggested that, perturbations in the exchange integral by small displacements of the $Se$ and $Ti$ orbitals can statically couple the electron and hole sublattices once a coherent excitonic condensate is formed, i.e. the  exciton-phonon mechanism. This effect is very similar to our proposal in that, they both arise due to the phenomenon of condensation and they additively cooperate in reducing the free energy. According to Ref.[\citen{Monney3}] the exciton-phonon coupling adds a negative contribution to the free energy which can be roughly written as $-({\cal F} g/\omega)^2$ where ${\cal F}$ is equivalent to our $\Delta_{{\bf k}}$ in Eq.(\ref{scs3}), $g$ is the exciton-phonon coupling strenght arising from the Fr\"{o}hlich type expansion of the exchange coupling $J$ and, $\omega$ is a typical phonon frequency. Our prediction for the static displacement due to the EC-force is similar to that in Ref.[\citen{Monney3}] which are comparable and nearly equal to the $\% 60$ of the experimental value\cite{Monney3}. Therefore attemps should be made to include both effects in an extended approach which may reflect the observations more realistically. In addition, the model presented in this work with two different mechanisms and two different orderings coupled self consistently is a plausible model that can explain the three distinct cases, i.e. $T_c^{CDW} < T_c^{EC}, T_c^{CDW} > T_c^{EC}$ and $T_c^{CDW}=T_c^{EC}$ (see the discussion below). 

\begin{figure}
\includegraphics[scale=0.7,angle=0]{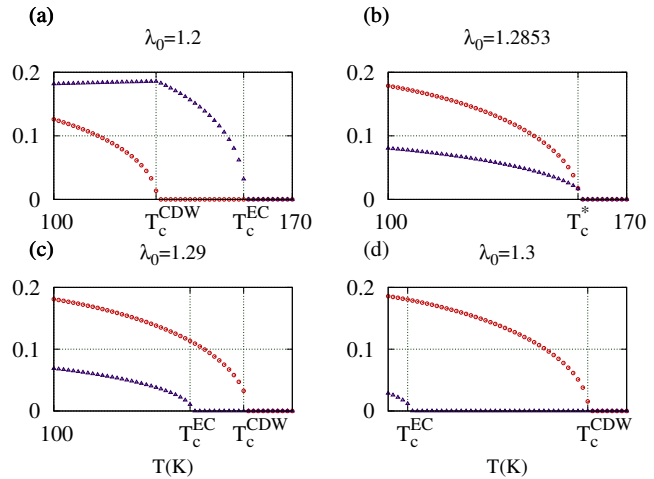}
\caption{(Color online) $ T_c^{EC} $ and $ T_c^{CDW} $ are illustrated for four different $ \lambda_0 $ values for $ n_0 \simeq 10^{14}cm^{-2} $ and $t_1=0$. a) $ T_c^{CDW} < T_c^{EC} $,  b) By increasing $ \lambda_0 $ the two critical temperatures were made to coincide at $ T=T^*_c $, c) After increasing $ \lambda_0 $ further, $T_c^{EC} < T_c^{CDW} $.   d) Increasing  $ \lambda_0 $ even further, the two $ T_c $'s can be widely separated.}  \vspace{-0.5cm}
\label{tc}
\end{figure}

Another result in this work concerns the general theory in Section.2. It is remarkable that the ratio $\chi=T_c^{EC}/T_c^{CDW} $ can be changed across unity, i.e. $\chi<1 \Rightarrow \chi >1$, by using different strengths of $ \lambda_{ep} $. The results are illustrated in Fig. \ref{tc} where we fixed in this case the exciton density $ n_0 \simeq 10^{14}cm^{-2} $. Here the $\lambda_{ep}$ is shown to play a sensitive role in the relative positions of $T_c^{CDW}$ and $T_c^{EC}$. This means that, there may be a variety of other TMDCs with stronger excitonic character where $ T_c^{EC} > T_c^{CDW} $. If there are such materials, this becomes an important result which can be pointing at that the nesting CDW does not play a major role in the lattice deformation.  
\vspace{0.3cm}

\section{Conclusions}
We firstly investigated in a general microscopic mean field approach, the competition between the CDW and the EC orders in a EHQW geometry. The phase diagram is covered by pure CDW, pure EC and coexistence regimes determined by the electron-phonon coupling and the quality of nesting. In the coexistence regime, a hybrid gap is opened which can persist even when the nesting is strongly suppressed. On the other hand, the dependence of the excitonic gap on the electron-hole separation shows the emergence of a new effect, i.e. a force arising due to the condensation. Combining this with the survival of the EC under weak nesting, we can conclude that the lattice instabilities and the crystal symmetries play a minor role in the EC, and hence the EC-force. Secondly, and encouraged by the observation that the crystal symmetries are expected to play a minor role in the EC-force, the results are applied to the two-dimensional TMDCs as typical systems for which the experimental data is known.

We used our general approach to explain the observed lattice deformation in these materials in the presence of the EC and propose that the EC-force is responsible for deforming the lattice. We compared the prediction of the theory with the experimental results known for $1 {\it T}$-$TiSe_2$. Although the CDW and the EC individually have two different mechanisms with different critical temperatures $T_c^{CDW}, T_c^{EC}$ the change in the lattice structure is due to the EC-force and occurs at the $T_c^{EC}$. We suggest that the nesting CDW does not play a major role in the structural phase transition. 

The model studied here predicts a full gap $\Delta_{hyb}$ in the electronic spectrum. At the first sight this would imply an insulating behaviour in all of the pure CDW, CDW/EC and pure EC phases. The first additional detail here that can change this interpretation is the presence of indirect electron-hole bands. Secondly, the existence of the semimetallic behaviour between these bands becomes important when applied to TMDCs. These exceptions are particularly important here since many of the TMDCs, and particularly $1 {\it T}$-$TiSe_2$, are believed to be semimetallic in both the normal and the low temperature ordered phases\cite{Fang_et.al}. Furthermore, the semimetallic character of the indirect electron and hole-like bands survives during the EC gap opening\cite{Li_et.al}. This behaviour, despite the opening of an energy gap in the spectrum was explained by Kohn as well as Halperin and Rice\cite{WK} based on Overhauser's CDW mechanism\cite{Overhauser}. Recent experiments on two dimensional $1 {\it T}$-$TiSe_2$ monolayers also show a similar semimetallic behaviour\cite{Sugawara}. Hence the presence of a finite hybrid gap $\Delta_{hyb}$ in the model studied here is not in conflict with the semimetallic nature of $1 {\it T}$-$TiSe_2$.    

More systematic experimental efforts are needed in these systems for examining the lattice distortion  which are likely to unravel the exciting phenomenon of the force arising due to condensation. The experimental discovery of the EC-force can lead to new exciting directions of theoretical and experimental condensed matter research as well as applications in new generation MEMS\cite{APL}. Recently, optomechanical cavities can provide an unsurpassible resource for the measurement of nanomechanical displacements. In this context, shifting the attention towards exciton-polariton condensation in such cavities may be crucial\cite{KLS}.  

\section*{Acknowledgement} The authors thank Vladimir Yudson for useful discussions.


\begin{thebibliography}{99}
\bibitem{10A} For our numerical calculations we are estimating the layer separation to be $10\AA$. Surely this is not the real separation between $Ti$ and $Se$ layers but this value is within the same order of magnitude with the real value which is $\sim 3.5\AA$. Moreover, the main outcomes of our numerical calculations, i.e. EC-force and it driving the periodic lattice distortions are insensitive to this estimation.
\end{thebibliography}
\end{document}